\documentclass[final,5p,times,twocolumn,preprint]{elsarticle}
\usepackage{slashed}
\usepackage{eucal} 
\usepackage{bm}
\usepackage{graphicx}
\usepackage{physics}
\usepackage[hypertexnames=false]{hyperref}
\usepackage{amsmath,amsfonts,amssymb,amscd,amsxtra,amsthm}
\allowdisplaybreaks 
\usepackage{color}
\begin{document}
\begin{frontmatter}
\title{Flavor structure of the energy-momentum tensor form factors of
  the proton}   
\author[i,l]{Ho-Yeon Won}
\ead{hoyeon.won@polytechnique.edu}

\author[i,k]{Hyun-Chul Kim}
\ead{hchkim@inha.ac.kr}
\author[j]{June-Young Kim}
\ead{jykim@jlab.org}
\address[i]{Department of Physics, Inha University, Incheon 402-751, South Korea} 
\address[l]{CPHT, CNRS, \'Ecole polytechnique, Institut Polytechnique de Paris,
91120 Palaiseau, France}
\address[k]{School of Physics, Korea Institute for Advanced Study 
   (KIAS), Seoul 02455, Republic of Korea}
\address[j]{Theory Center, Jefferson Lab, Newport News, VA 23606, USA}
\begin{abstract}
The energy-momentum tensor form factors furnish information on
the mechanics of the proton. It is essential to compute the
generalized isovector-vector form factors to examine the flavor
structure of the energy-momentum tensor form factors. The 
flavor-decomposed form factors reveal the internal structure of the proton.  The up
quark dominates over the down quark for the mass and spin of the
proton, whereas the down quark takes over the up quark for the
$D$-term form factor.  We investigate for the first time the isovector
$\bar{c}(t)$ form factor of the proton and its physical
implications. The flavor-decomposed $\bar{c}(t)$ form factors of the
proton unveil how the up-quark contribution is exactly canceled by 
the down-quark contribution inside a proton within the framework of
the pion mean-field approach. While the proton $\bar{c}(t)$ form factor
does not contribute to the proton mass, its flavor structure sheds
light on how the strong force fields due to the $\bar{c}(t)$ form
factor characterize the stability of the proton.  
\end{abstract}
\begin{keyword}
Energy-momentum tensor form factors of the proton, flavor
decomposition, $\bar{c}$ form factors, pion mean-field approach  
\end{keyword}
\end{frontmatter}


\section{Introduction}
The cosmological constant term (CCT) in Einstein's equation in general
relativity encodes the vacuum energy density of the universe, arising
from the quantum fluctuations~\cite{Einstein:1917ce,
  Zeldovich:1967gd, Weinberg:1988cp}.  The cosmological constant (CC)
is also known to be connected to the dark energy~\cite{Peebles:2002gy,
  Sola:2013gha}.  In nonperturbative quantum chromodynamics (QCD), the
gluon condensate gives the energy of the QCD
vacuum~\cite{Shifman:1978bx, Diakonov:2002fq}, which can be identified
as the QCD CC. If we decompose the proton matrix element of the
energy-momentum tensor (EMT) operator in terms of the EMT form factors,
one term is proportional to the metric tensor. Its coefficient is
called the proton $\bar{c}$ form factor, which has a similar structure
as the CCT in Einstein's equation, $\Lambda g_{\mu\nu}$. 
The proton $\bar{c}$ form factor furnishes critical information on
understanding the mechanics of the proton~\cite{Teryaev:2016edw, 
  Liu:2021gco}. Since the $\bar{c}$ form factors arise only 
when EMT current is not conserved, 
they are naturally scale-dependent~\cite{Hatta:2018sqd, Metz:2020vxd}.  
When both the quark and gluon degrees of freedom are considered, 
the proton $\bar{c}$ form factor disappears, because of conservation
of the EMT current.  However, the proton $\bar{c}$ form factor
comes into play when the flavor structure of the
proton energy-momentum tensor form factors (EMTFFs)   
as well as the proton mass decomposition is explored. The proton mass
is decomposed  in terms of $\pi N$ sigma terms, the quark and gluon energies, 
and the trace anomaly~\cite{Ji:1995sv, Ji:2021mtz}. On the other hand,
the proton $\bar{c}$ form factor was recently interpreted as the
isotropic pressure-volume work~\cite{Lorce:2017xzd,Lorce:2018egm} by
using the relation between the EMTFFs in the forward limit and 
the terminologies of perfect fluid in general relativity. 
It also gives a clue in understanding the partial internal energy 
inside a proton. When one investigates the flavor structure of 
the proton EMTFFs, the effects of the $\bar{c}$ form factor emerge. 
To carry out the flavor decomposition of the  EMTFFs, 
one has to compute the generalized isovector-vector form factors (GIVFFs). 
Since there is no physical reason for conservation of the isovector EMT-like current,
the isovector $\bar{c}$ form factor survives. 

In this Letter, we investigate the proton $\bar{c}$ form factors that
arise from the GIVFFs, which we currently have no empirical
information about. To calculate the proton EMTFFs and GIVFFs, 
we use the pion mean-field approach, also known as the chiral
quark-soliton model  ($\chi$QSM)~\cite{Diakonov:1987ty, Christov:1995vm}.  
The $\chi$QSM is built on an effective chiral action 
that is solely composed of the quark degrees of freedom. 
This effective chiral action is obtained 
by integrating out the gluon degrees of freedom 
from the instanton vacuum, and established in 
Refs.~\cite{Diakonov:1985eg,Diakonov:2002fq}. 
This process can be broken down into several steps: 
first, after the gluon degrees of freedom have been integrated out, 
the quark-quark interaction with a 2$N_f$ vertex is derived,
where $N_f$ denotes the number of flavors. 
This leads to the spontaneous breakdown of chiral symmetry, 
which results in the emergence of Nambu-Goldstone bosons and 
the dynamical quark mass. 
Next, the 2$N_f$ interaction is bosonized 
by incorporating pseudo-Nambu-Goldstone fields. 
Then we have integrated over the dressed quark fields 
to obtain the one-loop effective chiral action. 
The influence of gluons is effectively accounted for 
through the dynamical quark mass in a renormalization sense. 
Consequently, the quark EMT current alone is conserved 
in this framework, resulting in a vanishing proton $\bar{c}$ form factor 
from the quark EMT. Interestingly, the zero value of the proton
$\bar{c}$ form factor~\cite{Goeke:2007fp} is deeply linked to the von
Laue condition for proton stability. Ref.~\cite{Polyakov:2018exb}
showed that both quark and gluon contributions were insignificant.

The large $N_c$ behaviors of the $\bar{c}$ form factors in flavor
SU(2) symmetry are rather subtle. While the the isovector $\bar{c}$
form factor is proportional to $N_c^{-1}$, the isoscalar $\bar{c}$ form factor
is of $N_c^{0}$ order. On the other hand, the conservation of the EMT current forces 
the isoscalar $\bar{c}$ form factors to vanish. It imposes a strong
constraint on the isovector $\bar{c}$ form factor: The down-quark
component should always be the same as the negative up-quark component
in the present framework.

\section{Energy-momentum tensor form factors of the proton}
The matrix element of the bilocal quark and gluon vector operators on
the light cone are parametrized in terms of the vector GPDs
$H^{q,g}(x,\xi,t)$ and $E^{q,g}(x,\xi,t)$, where $q$ and $g$ denote the
quarks and gluon degrees of freedom, respectively. They are given as 
functions of the longitudinal momentum fraction carried by partons
$x$, the skewedness variable $\xi$, and the momentum 
transfer squared $t$. Here we consider quark contributions to them
only. In the leading-twist accuracy, this matrix element can be
expressed in terms of the unpolarized GPDs as
follows~\cite{Ji:1996nm}:   
\begin{align}
&   \hspace{-1.cm} \int   \frac{d\lambda}{2\pi}  e^{i\lambda x}  
    \mel{ p  ( p', J_{3}'  )   }
    {  \bar{\psi}_{q}  \left(  - \frac{\lambda n}{2}  \right)
      \slashed{n}   \psi_{q}  \left(  \frac{\lambda n}{2}  \right) }
      { p (  p,  J_{3} ) } \cr 
 &\hspace{-1.cm} = \bar{u} ( p',  J_{3}'  )  \Bigg[  H^{q} ( x,  \xi,
   t ) \slashed{n}  + E^{q} (x,  \xi,  t )   \frac{  i \sigma^{\mu\nu}
   n_{\mu} \Delta_{\nu}  }{2 M_{p} } \Bigg] u ( p, J_{3} ),
\end{align}
where $\psi_{q}$ is the quark field with flavor $q$ and  $M_{p}$
represents the proton mass. $p$ and $p'$ denote the initial and final
momenta. Their average and difference are defined by
$P=\left(p^{\prime}+p\right)/2$ and  $\Delta=p'-p$ with
$\Delta^{2}=t$, respectively. $n$ stands for a light-cone vector
satisfying $n\cdot\left(p^{\prime}+p\right)=2$.  
The longitudinal momentum fraction of a proton carried by a parton
is denoted by $x$ and the skewedness is expressed as $\xi$, 
which is defined as $n\cdot\Delta=-2\xi$. 
The first and second Mellin moments of  
the vector GPDs are identified as the electromagnetic (EM) form
factors and EMTFFs, respectively. Note that the Mellin moments of GPDs 
must satisfy the polynomiality, of which the maximal order is given as
$n+1$ due to Lorentz invariance~\cite{Ji:1996nm,  Ji:1996ek}. Thus,
the proton generalized form factors are defined by the $(n+1)$th
Mellin moments of the GPDs as follows~\cite{Diehl:2003ny}:    
\begin{align}
&   \hspace{-0.8cm} 
    \int_{-1}^{1} dx\;x^{n} 
    H^{q} ( x,  \xi,  t )  
  = \cr
&   \hspace{-0.8cm} 
    \sum_{i=0,\mathrm{even}}^{n} (2 \xi )^{i}
    A_{n+1 i}^{q} ( t )    
  + ( 2 \xi )^{n+1} C_{n+1 0}^{q} ( t )|_{n,\mathrm{odd}}  ,   \cr 
&   \hspace{-0.8cm} 
    \int_{-1}^{1} dx\;x^{n} 
    E^{q} ( x,  \xi,  t )    
  = \cr 
&   \hspace{-0.8cm} 
    \sum_{i=0,\mathrm{even}}^{n} ( 2 \xi )^{i}
    B_{n+1 i}^{q} ( t )   
  - ( 2 \xi )^{n+1} C_{n+1 0}^{q} ( t )|_{n,\mathrm{odd}},  
  \label{eq:2}
\end{align}
where $A_{n+1 i}^q$, $B_{n+1 i}^q$ and $C_{n+1 0}^q$ stand for the
generalized form factors of the quark part in QCD.  The first Mellin
moments $A_{1 0}^{q}(t)$ and $B_{1 0}^{q}(t)$ are identified as the
Dirac and Pauli form factors of the proton, $F^{q}_1(t)$ and
$F^{q}_2(t)$, respectively. 

The second Mellin moments are derived as 
\begin{align}
&   \int_{-1}^{1} dx  \, x H^{q}  ( x,  \xi,  t )  = A_{2 0}^{q}  (t)  
+ 4 C_{2 0}^{q}(t)  \xi^{2}  ,  \cr 
&   \int_{-1}^{1} dx  \, x E^{q}  ( x,  \xi,  t )  = B_{2 0}^{q}  (t)  
- 4 C_{2 0}^{q} (t) \xi^{2}  . 
\label{eq:3}
\end{align}
The EMTFFs are given by the linear combinations of the second Mellin
moments.  The matrix element of the symmetric EMT 
current
$\hat{T}^{\mu \nu, q}= \frac{1}{4}\bar{\psi}_{q} i
\overleftrightarrow{\mathcal{D}}^{ \{ \mu} \gamma^{\nu \} }
\psi_{q}$~\cite{Leader:2013jra, Lorce:2017wkb} with the 
covariant derivative $\overleftrightarrow{\mathcal{D}}^{\nu} =
\overleftrightarrow{\partial}^{\nu} -2 i g A^{\nu}$ 
and  
$\overleftrightarrow{\partial}^{\nu}=\overrightarrow{\partial}^{\nu} -
\overleftarrow{\partial}^{\nu}$ is parametrized in terms of 
the four different EMTFFs $A^{q}$, $J^{q}$, $D^{q}$, and $\bar{c}^{q}$:   
\begin{align}
&  \hspace{-0.6cm} 
    \mel{p(p',J_{3}')}   {   \hat{T}_{\mu\nu}^{q}  ( 0 )  }
                {p(p,J_{3})}\cr 
& \hspace{-0.6cm}= \bar{u}(p',J_{3}')\Bigg[
    A^{q}  ( t ) \frac{  P_{\mu} P_{\nu} }{  M_{p}}
    + J^{q}  ( t ) \frac{  i ( P_{\mu} \sigma_{\nu\rho} 
    + P_{\nu}    \sigma_{\mu\rho}  )\Delta^{\rho} } { 2M_{p}} 
                                   \cr
& + D^{q}  ( t )  \frac{ \Delta_{\mu} \Delta_{\nu} -
    g_{\mu\nu} \Delta^{2}}{4M_{p}}     
  + \bar{c}^{q} ( t ) M_{p} g_{\mu\nu}  \Bigg]  u(p,J_{3}),
  \label{eq:4}
\end{align}
where $A^q$, $J^q$, $D^q$, and $\bar{c}^q$ are called the
flavor-decomposed light-front~(LF) momentum, spin, $D$-term, and
$\bar{c}$ form factors, respectively. As mentioned above, the second
Mellin moments of the vector GPDs are related to the EMTFFs as follows 
\begin{align}
&   A_{2 0}^{q} ( t) =A^{q}  ( t ), \quad \frac{1}{2}  \Bigl[  A_{2
  0}^{q} ( t) + B_{2 0}^{q} ( t)  \Bigr] = J^{q}  ( t ) ,  \cr 
&   4 C_{2 0}^{q} ( t)= D^{q}  ( t ).
  \label{eq:5}
\end{align}
As we observe from Eqs.~\eqref{eq:3} and \eqref{eq:5}, the
leading-twist GPDs do not provide the proton $\bar{c}$ form factors.  
Higher-twist GPDs are required to define them~\cite{Leader:2012ar,
  Leader:2013jra}.  

Note that the symmetric EMT current is conserved only when both the
quark and gluon parts are considered: 
\begin{align}
\partial^{\mu}  \hat{T}_{\mu\nu}  = 0 , \qquad  \hat{T}_{\mu\nu}  =
  \hat{T}^{u+d}_{\mu\nu} + \hat{T}_{\mu\nu}^{g}. 
\label{eq:6}
\end{align}
At the zero momentum transfer $t$, thus, the EMTFFs $A^{u+d}$ and
$J^{u+d}$ (or proton EMTFFs) are normalized as $A^{p}=A^{u+d} ( 0 ) +
A^{g} ( 0 )  = 1$ and $J^{p}=J^{u+d} ( 0 ) + J^{g} ( 0 )  =
\frac{1}{2}$ together with the gluon contributions. However, there is
no such constraint on the GIVFFs as well as the $D$-term.   
Note that the GIVFFs are derived from the isovector EMT-like current
that is not conserved. The non-conserved isovector EMT-like current implies 
that $\bar{c}^q$ form factors with a specific flavor $q$ does not need
to vanish. Thus, the flavor-decomposed $\bar{c}$ form factors of the
proton should be finite. In the current work, we will scrutinize the
physical implications of the flavor-decomposed $\bar{c}^q$.

\section{Pion mean-field approach}
Since the $\chi$QSM has already been used for deriving the
EMTFFs~\cite{Goeke:2007fp, Kim:2020nug, Won:2022cyy}, we will mainly
present the main results for the EMTFFs and GIVFFs within the
framework of $\chi$QSM in flavor SU(2) symmetry.  
The $\chi$QSM is characterized by the low-energy QCD effective
partition function in Euclidean space~\cite{Diakonov:1985eg,
  Diakonov:1987ty, Christov:1995vm, Diakonov:2002fq}  
\begin{align}
  \hspace{-0.6cm}  \mathcal{Z}_{\mathrm{eff}}
&= \int  \mathcal{D} \pi^a \exp\left[  - S_{\mathrm{eff}}
  ( \pi^a )  \right], 
\end{align}
where $\pi^a$ is the pseudo-Nambu-Goldstone boson fields and 
$S_{\mathrm{eff}}$ denotes the effective chiral action expressed as  
\begin{align}
S_{\mathrm{eff}} = -N_c \mathrm{Tr}\log\left[
i\slashed{\partial} + i M e^{i\gamma_5 \pi^a \tau^a} + i\hat{m}\right]. 
\end{align}
$N_c$ designates the number of colors, $M$ denotes the dynamical quark
mass, and $\hat{m}$ is the current-quark mass matrix
$\mathrm{diag}(m_{\mathrm{u}},\,m_{\mathrm{d}})$. 
The Dirac Hamiltonian $h(U)$ is defined by
$h(U)=\gamma_{4}\gamma_{i}\partial_{i}
+\gamma_{4}Me^{i\gamma_{5}\pi^{a}\tau^{a}}+\gamma_{4}\bar{m}\mathbf{1}$ 
with the average value of the current-quark masses
$\bar{m}=\left(m_{u}+m_{d}\right)/2$. 
We assume isospin 
symmetry ($m_{\mathrm{u}}=m_{\mathrm{d}}$).  Introducing the hedgehog
ansatz $\pi^a = P(r) n^a$, we can determine the profile function
$P(r)$ by solving the classical equation of motion
self-consistently. Since the pion-loop corrections are of $1/N_c$, we
suppress them and carry out the functional integration over $\pi^a$,
considering the rotational and translational zero modes, which is
called the zero-mode quantization. Introducing the external tensor
source field, we can compute the matrix element of the EMT current. 

The proton matrix element of the symmetrized EMT current in Euclidean 
space can be calculated as follows:  
\begin{align}
&   \mel{p  (p',J_{3}')}    {\hat{T}_{\mu\nu}^{\chi}(0)} 
    {p (  p,J_{3})} \cr
& = \lim_{T\to\infty} \frac{1}{Z_{\mathrm{eff}}}
    \mathcal{N}^{*}(p')\mathcal{N}(p)
    e^{ip_{4}\frac{T}{2}-ip'_{4}\frac{T}{2}} \int d^{3}\bm{x} \,
    d^{3}\bm{y} \, e^{(-i\bm{p}' \cdot \bm{y} + i\bm{p} 
    \cdot \bm{x})}\cr 
&   \times 
    \int \mathcal{D}U \int \mathcal{D} \psi \mathcal{D}
    \psi^{\dagger}   
    J_{p}(\bm{y},T/2)    \hat{T}_{\mu\nu}^{\chi}(0)
    J_{p}^{\dagger}(\bm{x},-T/2)\cr
&   \times 
    \exp
    \left[ - S_{\mathrm{eff}}\right],
\end{align}
where $J_{p}$ represents the Ioffe-type current consisting of the
$N_{c}$ valence quarks~\cite{Ioffe:1981kw} and 
$\hat{T}_{\mu\nu}^{\chi}(0)$ denotes the
symmetrized EMT current derived from effective chiral theory 
in the Euclidean space. 
Note that the normalization factor $\mathcal{N}^{*}(p')\mathcal{N}(p')$ 
is reduced to the static normalization $2M_{p}$, and the proton state
implicitly carries the spin and isospin quantum numbers, i.e., 
$ J$, $J_{3}$, $T$, and $T_{3}$.  

The temporal, mixed, and spatial components of the EMT current are 
expressed as  
\begin{align}
    \hat{T}_{00}^{\chi} 
& = \frac{i}{2}
    \psi^{\dagger} 
    \left(
    \gamma_{4} \overrightarrow{\partial}_{4}
  - \gamma_{4} \overleftarrow{\partial}_{4} 
    \right) 
    \tau^{\chi}
    \psi, \cr
    \hat{T}_{0k}^{\chi} 
& = - \frac{1}{4}
    \psi^{\dagger} 
    \left(
    \gamma_{4} \overrightarrow{\partial}_{k}
  + \gamma_{k} \overrightarrow{\partial}_{4}
  - \gamma_{4} \overleftarrow{\partial}_{k} 
  - \gamma_{k} \overleftarrow{\partial}_{4} 
    \right) 
    \tau^{\chi}
    \psi, \cr
    \hat{T}_{ij}^{\chi} 
& = - \frac{i}{4}
    \psi^{\dagger} 
    \left(
    \gamma_{i} \overrightarrow{\partial}_{j}
  + \gamma_{j} \overrightarrow{\partial}_{i}
  - \gamma_{i} \overleftarrow{\partial}_{j} 
  - \gamma_{j} \overleftarrow{\partial}_{i} 
    \right) 
    \tau^{\chi}
    \psi,
\label{eq:EMT_current}
\end{align}
where we introduce the superscripts $\chi=0,3$ that represent
respectively the isoscalar $\chi=0 =u+d$ and isovector $\chi=3
=u-d$ components of the EMT current. 

Defining the static symmetric EMT distributions in a Wigner
sense~\cite{Polyakov:2002yz, Lorce:2020onh} 
\begin{align}
    T^{\chi}_{\mu\nu,p}  ( \bm{r} )
& = \int      \frac{d^{3}\Delta}{2M_{p}(2\pi)^{3}}  
    e^{-i\bm{\Delta}\cdot\bm{r}}    
    \mel{p (p^{\prime},J_{3}^{\prime})}    {\hat{T}_{\mu\nu}^{\chi}}
    {p\left(p,J_{3}\right)}  , 
\end{align}
we obtain the expressions for the flavor-decomposed EMTFFs in the large
$N_{c}$ limit.  The isoscalar components are expressed as
\begin{align}
    \left[
    A^{u+d}  ( t )     - \frac{t}{4M_{p}^{2}}  
    D^{u+d}  ( t )   
    \right] \delta_{J_{3}^{\prime}J_{3}} 
& = \frac{1}{M_{p}} \int d^{3}r \,
    j_{0}(r\sqrt{-t})
    \varepsilon^{u+d}_{p}  (r),\cr 
    \frac{t}{6M_{p}^{2}}  D^{u+d}(t)
    \delta_{J_{3}^{\prime}J_{3}}
& = \frac{1}{M_{p}}     \int d^{3}r \, 
    j_{0}(r\sqrt{-t})
    p^{u+d}_{p}(r),\cr 
    D^{u+d}(t)  \delta_{J_{3}^{\prime}J_{3}}
& = 4 M_N \int d^{3}r \, 
    \frac{j_{2}(r\sqrt{-t})}{t} 
    s_p^{u+d}(r),\cr
    2 S_{J_{3}^{\prime}J_{3}}^{3}
    J^{u+d}(t)  
& = 3 \int d^{3}r \, 
    \frac{j_{1}(r\sqrt{-t})}{r\sqrt{-t}} 
    \rho_{J,p}^{u+d} (r),
\end{align}
whereas the isovector components are written as 
\begin{align}
&   \left[
    A^{u-d}  ( t )   
  + \bar{c}^{u-d} ( t )    
  - \frac{t}{4M_{p}^{2}}  
    \left(  D^{u-d}  ( t )   -  2J^{u-d}  ( t ) \right)   
    \right] \delta_{J_{3}^{\prime}J_{3}}\cr
& = \frac{1}{M_{p}} \int d^{3}r \,
    j_{0}(r\sqrt{-t})
    \varepsilon^{u-d}_{p}  (r),\cr 
&   \left[
    \bar{c}^{u-d}(t) 
  - \frac{t}{6M_{p}^{2}}  D^{u-d}(t)
    \right] \delta_{J_{3}^{\prime}J_{3}}
  = - \frac{1}{M_{p}} 
    \int d^{3}r \, 
    j_{0}(r\sqrt{-t})   p_p^{u-d}(r),\cr
&   D^{u-d}(t)  \delta_{J_{3}^{\prime}J_{3}}
  = 4 M_p \int d^{3}r \, 
    \frac{j_{2}(r\sqrt{-t})}{t}     s_p^{u-d}(r),\cr
&   2 S_{J_{3}^{\prime}J_{3}}^{3}
    J^{u-d}(t)   = 3 \int d^{3}r \, 
    \frac{j_{1}(r\sqrt{-t})}{r\sqrt{-t}} 
    \rho_{J,p}^{u-d}(r),
\end{align}
where $\varepsilon_{p}$, $p_{p}$, $s_{p}$, and
$\rho_{J,p}$ denote the mass, pressure, shear force, and 
angular momentum distributions for the isoscalar and
isovector components, respectively. 
For the explicit expressions for these EMT distributions are given
in~\ref{app:A}. Note that they depend on the 
quantum numbers of the proton.  

Before we proceed to compute the EMTFFs, it is worthwhile to
mention about the polynomiality given in Eq.~\eqref{eq:2}. Since the
EMTFFs are regarded as the second Mellin moments, it is of great
importance to examine whether this polynomiality satisfies within the
framework of the $\chi$QSM. Noticeably, it was proven that the
polynomiality of the GPDs in Eq.~\eqref{eq:2} is preserved within the 
$\chi$QSM~\cite{Schweitzer:2002nm, Schweitzer:2003ms, Ossmann:2004bp}.    

Once we take the forward limit ($t\to0$) and
$J_{3}^{\prime}=J_{3}=1/2$, we get the flavor-decomposed mass, spin,
$D$-term, and $\bar{c}$ form factors as follows: For the    
isoscalar components, we have  
\begin{align}
&   A^{u+d}(0) = \frac{1}{M_{p}} \int d^{3}r  \,
    \varepsilon^{u+d}_{p}  \left(r\right) , \cr
&   D^{u+d}(0)=-\frac{4M_{p}}{15}\int d^{3}r \, r^{2} 
    s^{u+d}_{p}  \left(r\right),\cr 
&   \bar{c}^{u+d}(0)= -\frac{1}{M_{p}} \int d^{3}r \, 
    p^{u+d}_{p}  \left(r\right)=0,  \cr
&   J^{u+d}(0)=\int d^{3}r \, 
    \rho_{J,p}^{u+d} \left(r\right) ,
\label{eq:10}
\end{align}
whereas for the isovector components, we obtain 
\begin{align}
&   A^{u-d}(0) + \bar{c}^{u-d}(0) = \frac{1}{M_{p}} \int d^{3}r  \,
    \varepsilon^{u-d}_{p}  \left(r\right)  . \cr
&   D^{u-d}(0)=-\frac{4M_{p}}{15}\int d^{3}r \, r^{2} 
    s^{u-d}_{p}  \left(r\right),\cr 
&   \bar{c}^{u-d}(0)=-\frac{1}{M_{p}} \int d^{3}r \, 
    p^{u-d}_{p}  \left(r\right),  \cr
&   J^{u-d}(0)=\int d^{3}r \, 
    \rho_{J,p}^{u-d} \left(r\right).
\label{eq:10_1}
\end{align}
The first and third relations in Eq.~\eqref{eq:10} resemble 
the thermodynamic potentials for the partial internal energy and
isotropic pressure~\cite{Lorce:2017xzd,Lorce:2018egm}. 
Thus, the flavor-decomposed $\bar{c}(0)$ form factors contribute to
the decomposition  of the proton mass. 

Finally, we want to mention the $N_c$ counting of the EMTFFs and
GIVFFs, which are given as  
\begin{align}
&A^{u+d}(t) \sim O(N^{0}_{c}), \quad A^{u-d}(t) \sim O(N^{-1}_{c}), \cr
&J^{u+d}(t) \sim O(N^{0}_{c}), \quad J^{u-d}(t) \sim O(N^{1}_{c}), \cr
&D^{u+d}(t) \sim O(N^{2}_{c}), \quad D^{u-d}(t) \sim O(N^{1}_{c}), \cr
&\bar{c}^{u+d}(t) \sim O(N^{0}_{c}), \quad \bar{c}^{u-d}(t) \sim O(N^{-1}_{c}).
\end{align}
We observe that $A^{u+d}$ and $A^{u-d}$ have the same $N_c$ orders
as the isoscalar and isovector $\bar{c}$ form factors, respectively.

\section{Results and discussion}
Concerning the fixing of the parameters for the $\chi$QSM, we refer to 
Refs.~\cite{Christov:1995vm,Won:2022cyy}.

\begin{figure}[htp]
  \centering
  \includegraphics[scale=0.17]{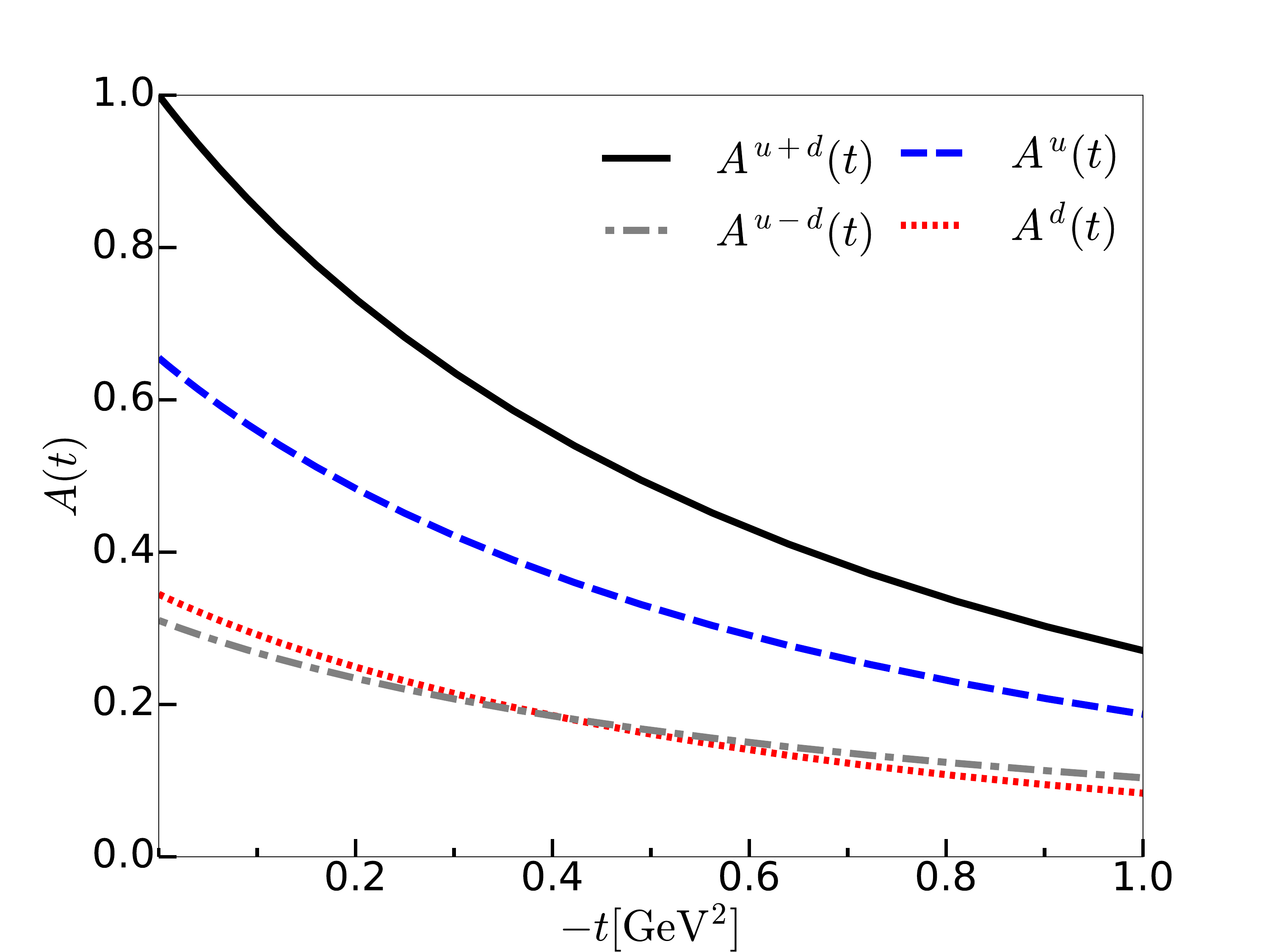}\\
  \includegraphics[scale=0.17]{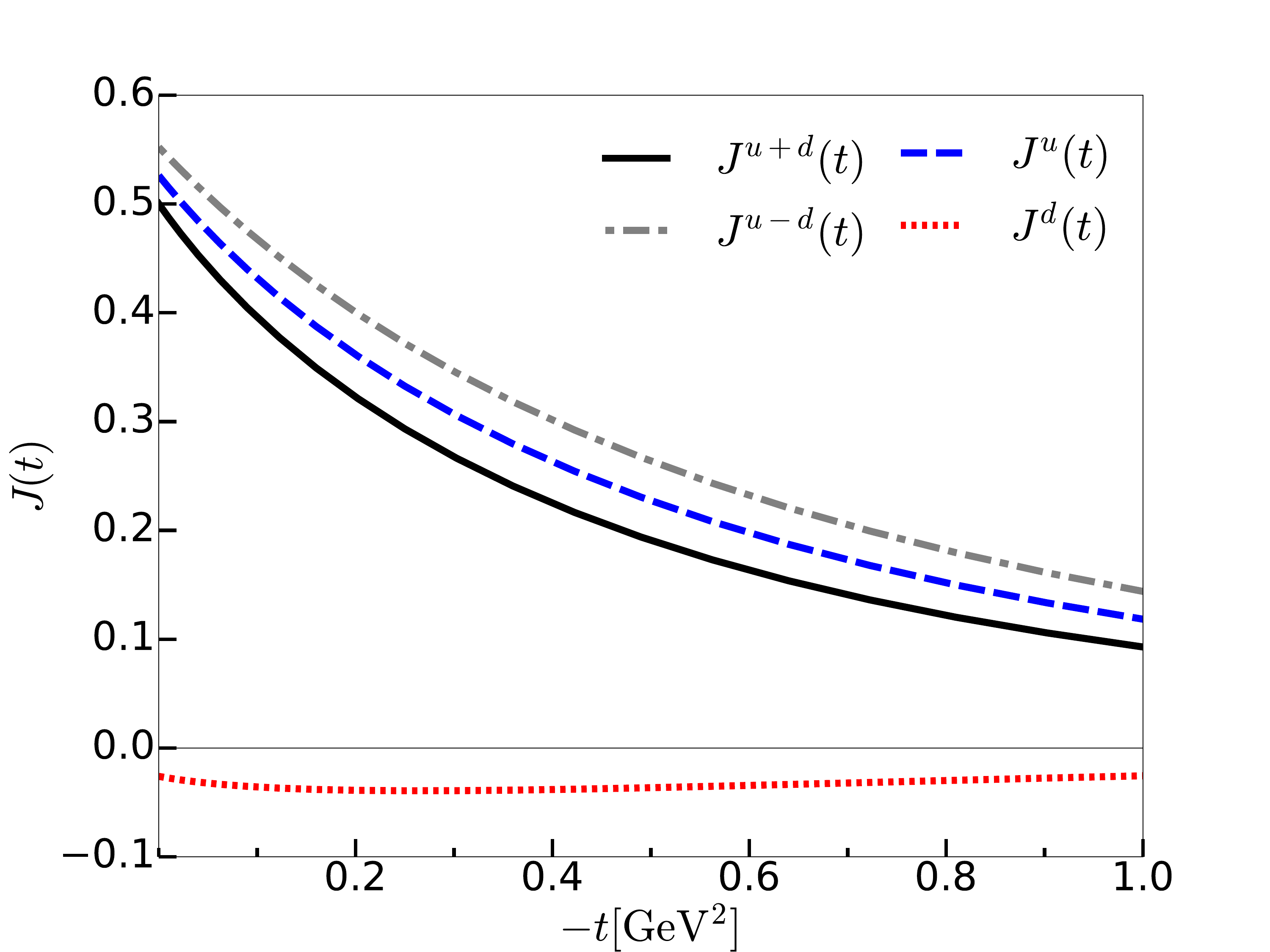}\\
  \includegraphics[scale=0.17]{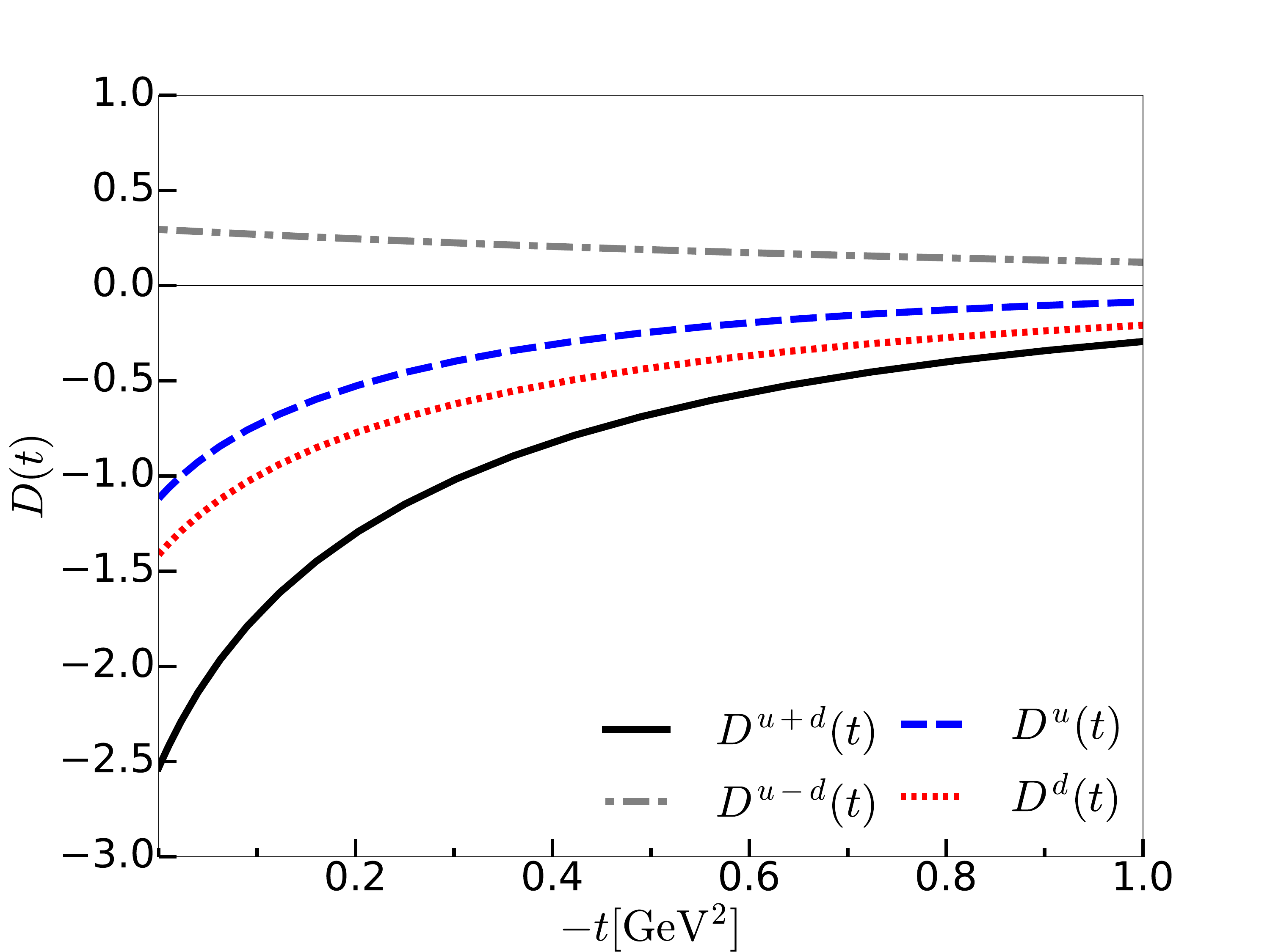}

  \caption{The flavor decompositions of the LF momentum, spin, and $D$-term
    form factors are drawn in the upper, middle, and lower panels,
    respectively.  The solid curves draw the corresponding
    EMTFFs and the dot-dashed ones depict the corresponding GIVFFs. The
    dashed and dotted ones represent the up-quark and down-quark
    contributions to the corresponding EMTFFs.   
}   
\label{fig:1}
\end{figure}
The solid curves in Fig.~\ref{fig:1} present the numerical results for
the EMTFFs, i.e. the $A(t)$ form factor, spin form factor, and the
$D$-term form factor in the upper, middle, and lower panels,
respectively.  The results are the same as those in
Refs.~\cite{Goeke:2007fp, Won:2022cyy, Wakamatsu:2007uc}.
The GIVFFs $A^{u-d}$, $J^{u-d}$, and $D^{u-d}$ can be considered as the isovector
partners corresponding to the EMTFFs. 
Decomposing the EMTFFs into the up-quark and down-quark form factors, 
we find a very interesting feature. 
The up quarks dominate over the down quarks for the LF momentum and
spin form factors. On the other hand, for the $D$-term form factor,
the down-quark contribution turns out slightly larger than the
up-quark one, both of which have negative values.  

\begin{figure}[htp]
  \centering
  \includegraphics[scale=0.17]{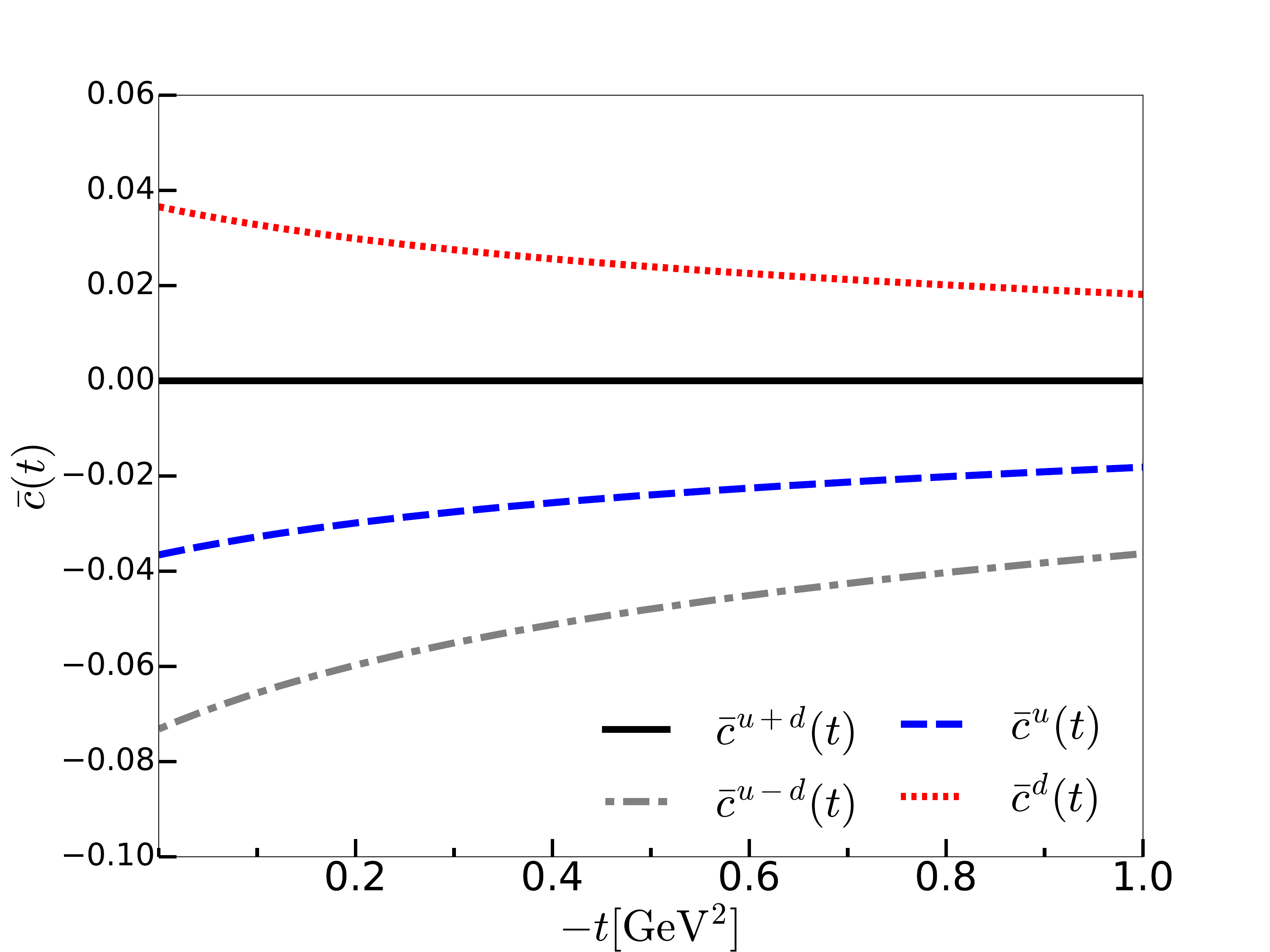}
  \caption{The flavor decomposition of the proton $\bar{c}$ form factor is
    drawn.  The solid curve draws proton $\bar{c}$ form factor and the
    dot-dashed ones depicts $\bar{c}^{u-d}$ . 
    The dashed and dotted ones represent the up-quark and down-quark
    contributions to the proton $\bar{c}$ form factor.
    }
\label{fig:2}
\end{figure}
Figure~\ref{fig:2} depicts the flavor decomposition 
of the proton $\bar{c}$ form factor. As
mentioned in the Introduction, the quark EMT 
current is conserved in the present work. As expected, thus,
$\bar{c}^{u+d}$ vanishes, which is also related to the Von Laue
condition. However, the $\bar{c}^{u-d}$ remains finite because the
isovector tensor current is in general not conserved. The solid and
dot-dashed curves in Fig.~\ref{fig:2} indicate this feature of the
proton $\bar{c}$ form factor. 
When we decompose it into the up- and down-quark contributions, 
their magnitudes are exactly the same but their signs are opposite each other, 
so that they are canceled each other. 
So, while the flavor-decomposed proton $\bar{c}$ form factors do not
contribute to the proton mass, they play a certain role in describing
the isotropic pressure-volume work inside a
proton~\cite{Lorce:2017xzd, Liu:2021gco}. 
In Table~\ref{tab:1}, we summarize the values of the flavor-decomposed
EMTFFs at zero momentum transfer, comparing them with the lattice
data~\cite{LHPC:2007blg, Gockeler:2003jfa}. Note that the
normalization scale of the $\chi$QSM is determined to be around $0.6$
GeV~\cite{Kim:1995bq}, whereas the lattice data~\cite{LHPC:2007blg,
  Gockeler:2003jfa} are derived at $\mu=2$ GeV.  
The present work and Refs.~\cite{Goeke:2007fp, Goeke:2007fq}
demonstrate that while the results for the isoscalar form factors are
comparable to those obtained from lattice calculations, that for the
$\bar{c}$ form factor is at variance with the lattice
results~\cite{Liu:2021gco}. When the $\chi$QSM was constructed from
the instanton vacuum, the gluon degrees of freedom were integrated out
via instantons. This means that we have only quark degrees of freedom
and the effects of the gluons are effectively absorbed in the
dynamical quark mass and pion mean fields. Thus we have correctly
obtained the zero value of the $\bar{c}$ form factor, which
effectively contains gluon contributions.  

We also found that the results for the isovector mass and angular
momentum form factors are in line with those from lattice
calculations. The comparison of the current results with the lattice
QCD requires an adjustment of the pion mass, and accordingly the
associated low-energy constants will be varied. In addition, the scale
evolution should be carried out. The comparison of the isoscalar
component has been done in Ref.~\cite{Goeke:2007fq}, and that of the
isovector component will appear elsewhere.   

The form factor $A^q$ in the forward limit is equivalent to the
momentum fraction of the proton: $A^q=\langle x\rangle_q$. 
While the total value of $A^q$ becomes $\sum_q A^q= \sum_{q}\langle
x\rangle_q=1$,  $A^q$ can not be identified as the flavor-decomposed
mass.  The reason can be found in the fact that each flavor component
of the $\bar{c}^{q}(0)$ form factors has a finite value, though the
total $\bar{c}$ form factor vanishes. Thus, the flavor-decomposed mass
can be expressed as   
\begin{align}
    M_p  = \sum_q M_{p}^{q}   = \sum_q 
    \left(A^q(0) + \bar{c}^q(0)
\right) M_{p}   
\label{eq:15}
\end{align}
which indicates $\sum_q \left(A^q(0)+ \bar{c}^q(0)\right)=1$.
Using Eq.~\eqref{eq:15}, 
we find the following inequalities between $M_{p}^{q}$ and $\expval{x}_{q}$: 
\begin{align}
    M_{p}^{u} / M_p
& = 62 \; \%     <
    \expval{x}_{u}
  = 66 \; \%  \cr
   M_{p}^{d} / M_p
  & = 38 \; \% 
    >
    \expval{x}_{d}
  = 34 \; \%  .
\end{align}
The flavor-decomposed mass differs from the corresponding component of
the momentum fraction by about $4~\%$ for both the up- and
down-quark contributions. 
In addition, we obtain the following relations: 
if $\bar{c}^{q} (0)>0$, then $M_{p}^{q}/M_{p}>\expval{x}_{q}$. 
If the flavor-decomposed $\bar{c}^{q}(0)$ form factor is zero, 
then $M_{p}^{q}/M_{p}=\expval{x}_{q}$.  

Note that the isovector component has a much weaker scale dependence
compared to the isoscalar or individual quark flavor components, 
since there is no gluon contribution to the isovector component. 
So, it implies that even though we consider the scale dependence 
of $\bar{c}$ form factor the results would not be much changed. 
Regarding the $D$-terms, we observed a significant difference between
our predicted $D$-term form factor and the lattice result. 
However, the numerical uncertainties associated with both the isovector 
and isoscalar $D$-term form factors obtained from lattice QCD are substantial. 
These uncertainties can even affect the sign of the form factors. 
\begin{table*}[!]
\caption{We list the flavor-decomposed proton EMTFFs at $t=0$ and
  compare them with results from lattice
  QCD~\cite{LHPC:2007blg,Gockeler:2003jfa}, which 
  are obtained from the chiral extrapolation at $t=0$ and
  $m_{\pi,\mathrm{phys}}$.}   
\begin{center}
  \renewcommand{\arraystretch}{1.4}
\scalebox{0.9}{%
\begin{tabular}{ccccc} 
  \hline
  \hline
  &
  & This work ($\mu\approx 0.6~\mathrm{GeV}$)  
  & Lattice QCD ($\mu=2~\mathrm{GeV}$)~\cite{LHPC:2007blg} 
  & Lattice QCD ($\mu=2~\mathrm{GeV}$)~\cite{Gockeler:2003jfa}\\
  \hline
  &   $A^{u}  ( 0 ) $
  &   $0.66$
  &   $0.34$  
  &   $0.40$  \\
  &   $A^{d} ( 0 ) $  
  &   $0.34$
  &   $0.18$
  &   $0.15$\\
  &   $J^{u}  ( 0 ) $  
  &   $0.53$  
  &   $0.21$  
  &   $0.37$  \\
  &   $J^{d} ( 0 ) $
  &  -$0.03$
  &  -$0.00$
  &  -$0.04$\\
  &  $D^{u}  ( 0 ) $
  & -$1.12$
  & -$0.57$  
  & -$0.54$  \\
  & $D^{d} ( 0 ) $         
  &  -$1.41$
  &  -$0.50$ 
  &  -$0.28$ \\
  & $\bar{c}^{u}  ( 0 ) $
  & -$0.04$
  & $-$  
  & $-$  \\
  & $\bar{c}^{d} ( 0 ) $         
  &  $0.04$
  &  $-$ 
  &  $-$ \\
  \hline
  \hline
\end{tabular}}
\end{center}
\label{tab:1}
\end{table*}

\begin{figure}[!]
  \centering
  \includegraphics[scale=0.17]{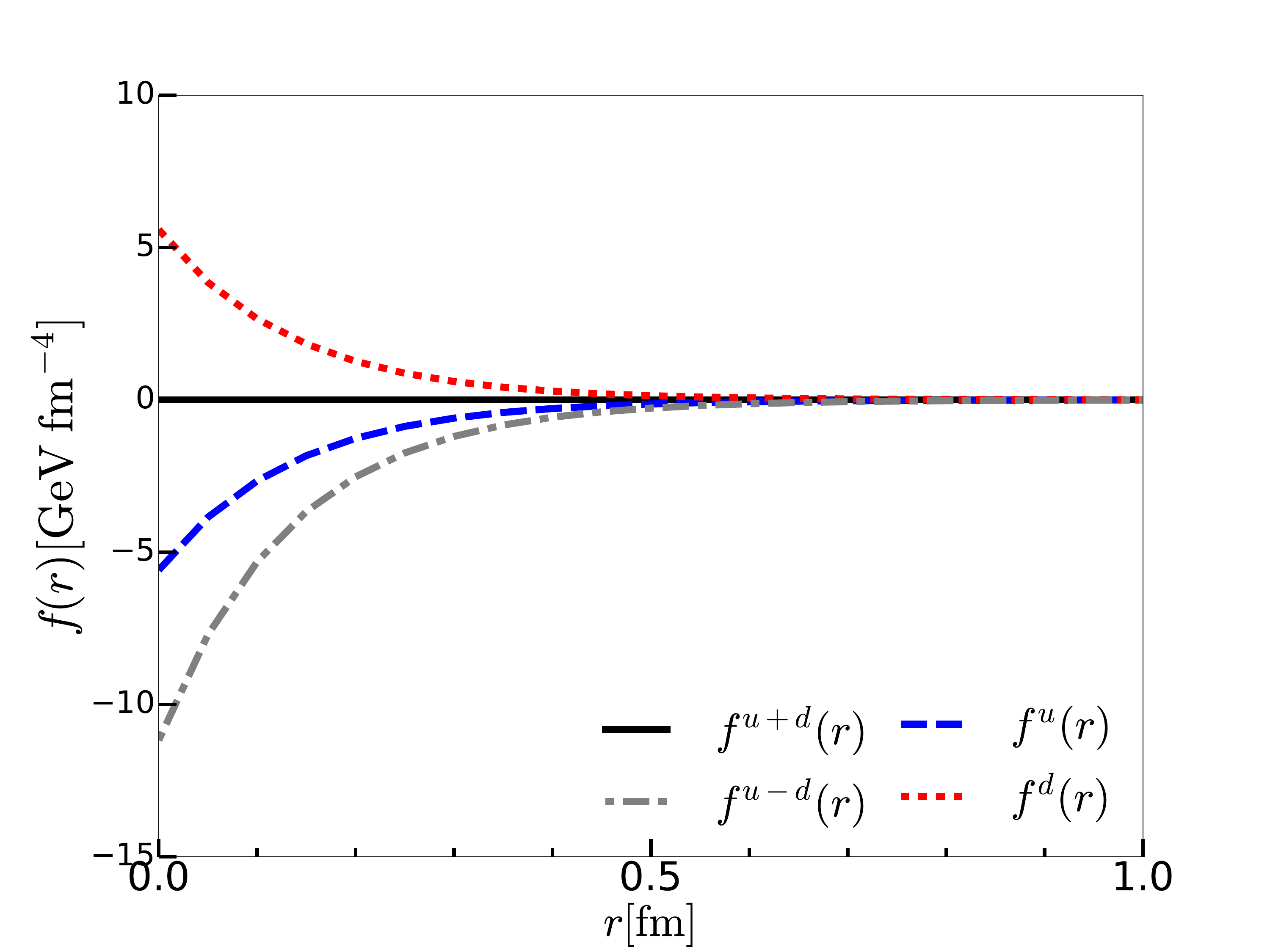}
  \caption{Contribution of the $\bar{c}$ form factors to the
    internal force fields inside a proton. Notations are the same as
    in Fig.~\ref{fig:2}. 
}
\label{fig:3}
\end{figure}
As mentioned previously, the proton $\bar{c}$ form factor contributes to
the isotropic pressure-volume work and force fields. 
As secured by the global stability condition for the
proton~\cite{Polyakov:2018zvc}, the pressure distribution of the
proton is balanced between the level-quark 
and Dirac-continuum (pion-cloud) contributions~\cite{Goeke:2007fp}  
in the $\chi$QSM. When it comes to the proton $\bar{c}$ form factor, 
the cancelation takes place between the up- and down-quark contributions. 
To demonstrate it, 
we first derive the contributions of the flavor-decomposed $\bar{c}$
form factors by the Fourier transforms. 
In doing so, we parametrize $\bar{c}^{u-d}$
in terms of the dipole-type parametrization 
with the parameters $\Lambda^{u-d}=1.5$ GeV.  
The conservation of the EMT current yields $\partial^{i}
T_{ij}^{u+d} = f_{j}^{u}+ f_{j}^{d} =0 $, 
which can be considered as an equilibrium equation 
for the internal forces between the $u$- and $d$-quark subsystems. 
A similar interpretation of the internal forces 
between the quark and gluon subsystem was conducted in
Ref.~\cite{Polyakov:2018exb}. 
The force-field vectors $f_{j}^{q}$ and their magnitudes $f^q$
are derived from the $\bar{c}$ form factors as follows:
\begin{align}
    f_{j}^{q} 
& = - M_{p} \frac{\partial}{\partial r^{j}} \int
    \frac{d^{3}\Delta}{(2\pi)^{3}} \;e^{- i \bm{\Delta} \cdot \bm{r}}
    \bar{c}^{q}(t),\cr 
    f^{q} 
& = - M_{p} \frac{\partial}{\partial r} \int
    \frac{d^{3}\Delta}{(2\pi)^{3}} \;e^{- i \bm{\Delta} \cdot \bm{r}}
    \bar{c}^{q}(t),
\end{align} 
where the spherical symmetry is imposed as $f_{j}^{q}=\hat{n}_{j}f^{q}$.
In Fig.~\ref{fig:3}, we illustrate $f^q(r)$. As already shown in
Fig.~\ref{fig:2},  $f^u$ is exactly canceled by $f^d$. 
Interestingly, the up-quark force field directs toward the
center of the proton, whereas the down-quark on pushes outward. 
As a result, the force field from the $\bar{c}$ form factor vanishes. 

\begin{figure}[htp]
  \centering
  \includegraphics[scale=0.55]{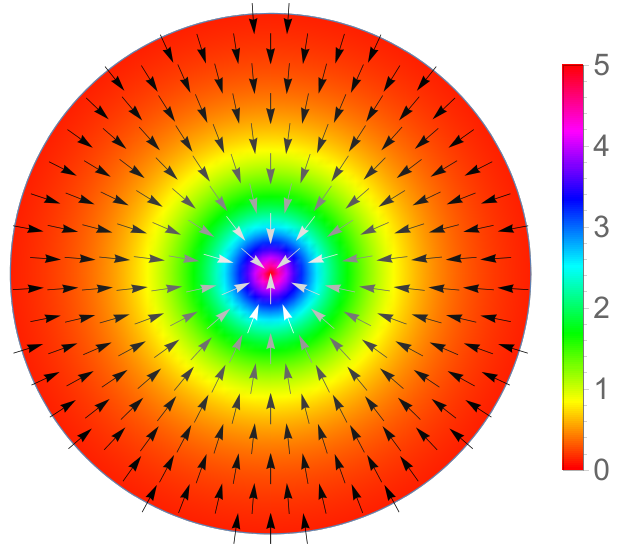}
  \includegraphics[scale=0.55]{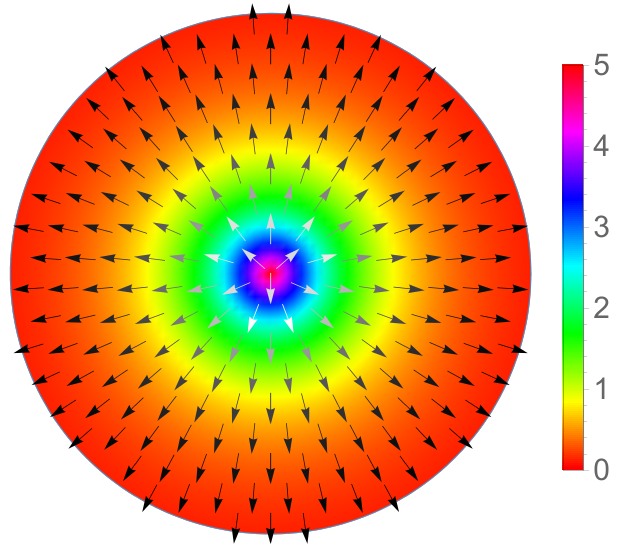}
  \caption{Visualization of the flavor-decomposed force fields from
    the proton $\bar{c}$ form factor. In the upper (lower) panel, 
    the up-quark (down-quark) force field is illustrated.}
\label{fig:4}
\end{figure}

In Fig.~\ref{fig:4} we visualize the flavor-decomposed force fields
from the proton $\bar{c}$ form factor, which portrays how $f^u (r)$
and $f^d(r)$ are distributed inside a proton. They are canceled each
other at each point,  so that the effects of the $\bar{c}$ form factor
completely vanish for the proton.

\section{Conclusions}
The general isovector-vector form factors of the proton enable us to
perform the flavor decomposition  of the proton energy-momentum tensor
form factors, which reveal novel features for the mechanical properties of
the proton. While the up-quark contributions dominate the proton
light-front momentum and spin form factors, the up- and down-quark
contributions to the $D$ term form factor are rather well
balanced. However, in contrast to the light-front momentum and spin
form factors, the magnitude of the down-quark contribution is slightly
larger than that of the up-quark contribution.  The proton isovector 
$\bar{c}$ form factor arises from the nonconservation of the isovector
energy-momentum tensor-like current. This yields the flavor-decomposed
the $\bar{c}$ form factors that do not vanish. They exhibit partial
internal pressure and energy inside a proton, which are canceled each other. 
It results in vanishing the proton $\bar{c}$ form factor. In conclusion, 
the flavor-decomposed $\bar{c}$ form factors shed light on how the
quarks describe the mechanics in the proton. 

\section*{Acknowledgements}
Authors want to express gratitude to Christian J. Weiss, Jose
L. Goity, C\'{e}dric Lorc\'{e}, and Keh-Fei Liu for the invaluable 
comments and discussions.  We are also grateful to Hyeon-Dong Son
for the discussion about the gluon contribution to the proton
$\bar{c}$ form factors. The work was supported by the Basic Science 
Research Program through the National Research Foundation of Korea
funded by the Korean government (Ministry of Education, Science and
Technology, MEST), Grant-No. 2021R1A2C2093368 and
2018R1A5A1025563. This work is also supported by the
U.S. Department of Energy, Office of Science, Office of Nuclear
Physics under contract DE-AC05-06OR23177 (JYK).
HYW is supported by the France Excellence scholarship through Campus France
funded by the French government (Ministère de l’Europe et des Affaires
Étrangères), 141295X.

\clearpage
\appendix
\section{Expressions of the 3D EMT distributions 
in the $\chi$QSM \label{app:A}}  
In this \ref{app:A}, we compile the explicit expressions of the 3D
EMT distributions. The Dirac Hamiltonian is diagonalized by the
corresponding eigenenergies and eigenfunctions
\begin{align}
    h(U)\psi_{n}(r) 
  = E_{n} \psi(r),
\end{align}
where $E_{n}$ and $\psi_{n}$ stand for the eigenenergies and
eigenfunctions of the Dirac Hamiltonian $h(U)$, respectively.
In the $\chi$QSM, the explicit expressions for the distributions are
provided by 
\begin{align}
    \varepsilon_{p}^{\chi}  (r)  
& = \mathcal{E}(r)
    \delta^{\chi 0}
  - \frac{2}{I_{1}}
    \expval{D_{\chi i}J_{i}}_{p} 
    \mathcal{J}_{1}(r)
    \delta^{\chi 3},\cr
    \rho_{J,p}^{\chi} (r)  
& = \expval{D_{\chi 3}}_{p} 
    \left(  
    \mathcal{Q}_{0} (r) 
  + \frac{1}{I_{1}} 
    \mathcal{Q}_{1} (r) 
    \right) 
    \delta^{\chi 3}
  - \expval{J_{3}}_{p}
    \frac{1}{I_{1}}     
    \mathcal{I}_{1} (r)
    \delta^{\chi 0} , \cr
    s_{p}^{\chi}  (r)
& = \mathcal{N}_{1}(r)
    \delta^{\chi 0}
  - \frac{2}{I_{1}}
    \expval{D_{\chi i}J_{i}}_{p}
    \mathcal{J}_{3}(r)
    \delta^{\chi 3},\cr
    p_{p}^{\chi}  (r)
& = \mathcal{N}_{3}(r)
    \delta^{\chi 0}
  - \frac{2}{I_{1}}
    \expval{D_{\chi i}J_{i}}_{p} 
    \mathcal{J}_{5}(r)
    \delta^{\chi 3},
\end{align}
where $I_{1}$ is the moment of intertia (see
Ref.~\cite{Christov:1995vm}). The $\expval{\cdots}_{p}$ represents   
the matrix element of the collectibe operators of the proton as
follow: 
\begin{align} 
    \expval{\cdots}_{p}
  = \int dR \, 
    \Psi^{*}_{(T'T'_{3})(J'J'_{3})} (R) \, 
    ... \, 
    \Psi^{}_{(TT_{3})(J J_{3})} (R).
\end{align}
The collective wave function of the proton $\Psi^{}_{(TT_{3})(JJ_{3})}
(R)$ is given by the SU(2) Wigner $D$ function 
\begin{align}
\Psi^{}_{(TT_{3})(J J_{3})} (R) :=\sqrt{2T+1} (-1)^{T+T_{3}}
  D^{J=T}_{-T_{3},J_{3}}(R). 
\end{align} 
The distributions for the LF momentum form factors are expressed as  
\begin{align}
    \mathcal{E} (r )   
& = N_{c}
    \left[ 
    E_{v}
    \psi_{v}^{\dagger}  (r  ) \psi_{v}  (r  )
  + \sum_{n} \psi_{n}^{\dagger}  (r  ) \psi_{n}  (r  ) R_{0n} 
    \right], \cr
    \mathcal{J}_{1} (r  )
& = \frac{N_{c}}{4} \Bigg[  \sum_{n\neq v}  \frac{E_{n}+E_{v}}{E_{n}-E_{v}}
    \mel{n}{\tau_{3}}{v}  \psi_{v}^{\dagger}  (r  ) \tau_{3}  \psi_{n}  (r  ) \cr
& + \frac{1}{2} \sum_{n,m} ( E_{n} + E_{m} )
    \mel{n}{\tau_{3}}{m}  \psi_{m}^{\dagger}  (r  ) \tau_{3}  \psi_{n}
  (r  ) R_{3nm}  \Bigg],   
\end{align}
those for the spin form factors are given as 
\begin{align}
    \mathcal{Q}_{0} (r  ) 
& = \frac{N_{c}}{4} \Bigg[ \psi_{v}^{\dagger}  (r  ) \Gamma^{J}_{vv3}
  \tau_{3}  \psi_{v}  (r  )   \cr 
& - \frac{1}{2} \sum_{n} \mathrm{sign} (E_{n}) \psi_{n}^{\dagger}  (r
  ) \Gamma^{J}_{nn3} \tau_{3}  \psi_{n}  (r  )  \Bigg]  , \cr 
    \mathcal{Q}_{1} (r  ) 
& = \frac{N_{c}}{4} i \epsilon_{ij3} 
    \Bigg[  \sum_{n\neq v}  \frac{\mathrm{sign}(E_{n})}{E_{n}-E_{v}}
  \mel{n}{\tau_{i}}{v} 
    \psi_{v}^{\dagger}  (r  ) \tau_{j}  \Gamma^{J}_{vn3} \psi_{n}  (r
  ) \cr 
& + \frac{1}{2} \sum_{n,m} \mel{n}{\tau_{i}}{m}
    \psi_{m}^{\dagger}  (r  ) \tau_{j}  \Gamma^{J}_{mn3} \psi_{n}  (r
  ) R_{6nm}  \Bigg], \cr 
    \mathcal{I}_{1} (r) 
& = \frac{1}{4} 
    \sum_{n\neq v}  
    \frac{1}{E_{n}-E_{v}} 
    \mel{n}{\tau_{3}}{v}
    \psi_{v}^{\dagger}  (r) 
    \Gamma_{nv3}^{J}
    \psi_{n}  (r)   \cr
& + \frac{1}{8} 
    \sum_{\substack{n=\mathrm{all}\\m=\mathrm{all}}} 
    \mel{n}{\tau_{3}}{m}  
    \psi_{m}^{\dagger}  (r) 
    \Gamma_{mn3}^{J}
    \psi_{n}  (r) 
    R_{3}(E_{n},E_{m}),   \cr
\end{align}
those for the $D$-term form factors and $\bar{c}$ form factors are
respectively written as 
\begin{align}
    \mathcal{N}_{1}(r)
& = \frac{3}{2} N_{c}\Bigg[\psi_{v}^{\dagger}(r)\Gamma^{s}\psi_{v}(r) 
  + \sum_{n}\psi_{n}^{\dagger}(r)\Gamma^{s}\psi_{n}(r)R_{1n}\Bigg]  , \cr
    \mathcal{J}_{3}(r)
& = \frac{3}{4} N_{c} \Bigg[\sum_{n\neq v}\frac{\mel{n}{\tau_{3}}{v}}{E_{n}-E_{v}}
    \psi_{v}^{\dagger}(r)\tau_{3}\Gamma^{s}
    \psi_{n}(r)  \cr
& + \frac{1}{2}\sum_{n,m}\mel{n}{\tau_{3}}{m}
    \psi_{m}^{\dagger}(r)\tau_{3}\Gamma^{s}
    \psi_{n}(r)R_{5nm} \Bigg],  
\end{align}
and 
\begin{align}
    \mathcal{N}_{3}(r) 
& = \frac{N_{c}}{3}\Bigg[\psi_{v}^{\dagger}(r)
    \Gamma^{p}\psi_{v}(r)
  + \sum_{n}\psi_{n}^{\dagger}(r)\Gamma^{p}\psi_{n}(r)R_{1n}\Bigg]  , \cr
    \mathcal{J}_{5}(r)
& = \frac{N_{c}}{6}\Bigg[\sum_{n\neq v}\frac{\mel{n}{\tau_{3}}{v}}{E_{n}-E_{v}}
    \psi_{v}^{\dagger}(r)\tau_{3}
    \Gamma^{p}
    \psi_{n}(r) \cr
& + \frac{1}{2}\sum_{n,m}\mel{n}{\tau_{3}}{m}
    \psi_{m}^{\dagger}(r)\tau_{3}\Gamma^{p}
    \psi_{n}(r)R_{5nm}\Bigg].  
\end{align}
Note that $\psi_{v}(r):=\langle r| v \rangle$ and
$\psi_{n}(r):=\langle r| n \rangle$. The total angular momentum
$\Gamma^{J}_{mn3}$, shear 
force $\Gamma^{s}$, and pressure $\Gamma^{p}$ operators are
respectively expressed by 
\begin{align}
    \Gamma^{J}_{nm3} 
  &= \left[  
    2 \hat{L}_{3} 
  + \left( E_{n} + E_{m} \right) \gamma_{5}  
    (r  \times  \bm{\sigma}  )_{3} \right], \cr
    \Gamma^{s} 
& = \gamma^{0}\left(\bm{\hat{n}}\cdot\bm{p}\right) 
  - \frac{1}{3}\gamma^{0}\left(\bm{\gamma}\cdot\bm{p}\right), \cr
    \Gamma^{p} 
& = \gamma^{0}(\bm{\gamma}\cdot\bm{p}),
\end{align}
where the orbital angular momentum is defined by  
\begin{align}
\hat{\bm{L}}
  = \left[r \times \frac{i}{2} 
    \left( \overleftarrow{\bm{\nabla}}
  - \overrightarrow{\bm{\nabla}}\right)  \right].  
\end{align}

The regularization functions for the distributions are expressed as  
\begin{align}
  R_{0n} & = \frac{1}{4\sqrt{\pi}} \int_{\Lambda^{-2}}  \frac{du}{u^{3/2}}  e^{-uE_{n}^{2}}, \cr
  R_{1n} & = - \frac{E_{n}}{2\sqrt{\pi}}\int_{\Lambda^{-2}}  \frac{du}{\sqrt{u}} e^{-uE_{n}^{2}},\cr
  R_{3nm} & = \frac{1}{2\sqrt{\pi}}\int_{\Lambda^{-2}} \frac{du}{\sqrt{u}} 
  \bigg[\frac{1}{u}\frac{e^{-uE_{n}^{2}}-e^{-uE_{m}^{2}}}{E_{m}^{2}-E_{n}^{2}}  \cr
& - \frac{E_{n}e^{-uE_{n}^{2}}+E_{m}e^{-uE_{m}^{2}}}{E_{n}+E_{m}}\bigg],\cr
  R_{5nm} & =  \frac{1}{2}\frac{\mathrm{sign}(E_{n}) -
            \mathrm{sign}(E_{m})}{E_{n}-E_{m}},\cr  
  R_{6nm} & = \frac{1-\mathrm{sign}(E_{n})
            \mathrm{sign}(E_{m})}{E_{n}-E_{m}} .
\end{align}
\bibliography{NCC}

\begin{thebibliography}{10}
\expandafter\ifx\csname url\endcsname\relax
  \def\url#1{\texttt{#1}}\fi
\expandafter\ifx\csname urlprefix\endcsname\relax\def\urlprefix{URL }\fi
\expandafter\ifx\csname href\endcsname\relax
  \def\href#1#2{#2} \def\path#1{#1}\fi

\bibitem{Einstein:1917ce}
A.~Einstein, {Cosmological Considerations in the General Theory of Relativity},
  Sitzungsber. Preuss. Akad. Wiss. Berlin (Math. Phys. ) 1917 (1917) 142--152.

\bibitem{Zeldovich:1967gd}
Y.~B. Zeldovich, {Cosmological Constant and Elementary Particles}, JETP Lett. 6
  (1967) 316.

\bibitem{Weinberg:1988cp}
S.~Weinberg, {The Cosmological Constant Problem}, Rev. Mod. Phys. 61 (1989)
  1--23.
\newblock \href {https://doi.org/10.1103/RevModPhys.61.1}
  {\path{doi:10.1103/RevModPhys.61.1}}.

\bibitem{Peebles:2002gy}
P.~J.~E. Peebles, B.~Ratra, {The Cosmological Constant and Dark Energy}, Rev.
  Mod. Phys. 75 (2003) 559--606.
\newblock \href {http://arxiv.org/abs/astro-ph/0207347}
  {\path{arXiv:astro-ph/0207347}}, \href
  {https://doi.org/10.1103/RevModPhys.75.559}
  {\path{doi:10.1103/RevModPhys.75.559}}.

\bibitem{Sola:2013gha}
J.~Sola, {Cosmological constant and vacuum energy: old and new ideas}, J. Phys.
  Conf. Ser. 453 (2013) 012015.
\newblock \href {http://arxiv.org/abs/1306.1527} {\path{arXiv:1306.1527}},
  \href {https://doi.org/10.1088/1742-6596/453/1/012015}
  {\path{doi:10.1088/1742-6596/453/1/012015}}.

\bibitem{Shifman:1978bx}
M.~A. Shifman, A.~I. Vainshtein, V.~I. Zakharov, {QCD and Resonance Physics.
  Theoretical Foundations}, Nucl. Phys. B 147 (1979) 385--447.
\newblock \href {https://doi.org/10.1016/0550-3213(79)90022-1}
  {\path{doi:10.1016/0550-3213(79)90022-1}}.

\bibitem{Diakonov:2002fq}
D.~Diakonov, {Instantons at work}, Prog. Part. Nucl. Phys. 51 (2003) 173--222.
\newblock \href {http://arxiv.org/abs/hep-ph/0212026}
  {\path{arXiv:hep-ph/0212026}}, \href
  {https://doi.org/10.1016/S0146-6410(03)90014-7}
  {\path{doi:10.1016/S0146-6410(03)90014-7}}.

\bibitem{Teryaev:2016edw}
O.~V. Teryaev, {Gravitational form factors and nucleon spin structure}, Front.
  Phys. (Beijing) 11~(5) (2016) 111207.
\newblock \href {https://doi.org/10.1007/s11467-016-0573-6}
  {\path{doi:10.1007/s11467-016-0573-6}}.

\bibitem{Liu:2021gco}
K.-F. Liu, {Proton mass decomposition and hadron cosmological constant}, Phys.
  Rev. D 104~(7) (2021) 076010.
\newblock \href {http://arxiv.org/abs/2103.15768} {\path{arXiv:2103.15768}},
  \href {https://doi.org/10.1103/PhysRevD.104.076010}
  {\path{doi:10.1103/PhysRevD.104.076010}}.

\bibitem{Hatta:2018sqd}
Y.~Hatta, A.~Rajan, K.~Tanaka, {Quark and gluon contributions to the QCD trace
  anomaly}, JHEP 12 (2018) 008.
\newblock \href {http://arxiv.org/abs/1810.05116} {\path{arXiv:1810.05116}},
  \href {https://doi.org/10.1007/JHEP12(2018)008}
  {\path{doi:10.1007/JHEP12(2018)008}}.

\bibitem{Metz:2020vxd}
A.~Metz, B.~Pasquini, S.~Rodini, {Revisiting the proton mass decomposition},
  Phys. Rev. D 102 (2020) 114042.
\newblock \href {http://arxiv.org/abs/2006.11171} {\path{arXiv:2006.11171}},
  \href {https://doi.org/10.1103/PhysRevD.102.114042}
  {\path{doi:10.1103/PhysRevD.102.114042}}.

\bibitem{Ji:1995sv}
X.-D. Ji, {Breakup of hadron masses and energy - momentum tensor of QCD}, Phys.
  Rev. D 52 (1995) 271--281.
\newblock \href {http://arxiv.org/abs/hep-ph/9502213}
  {\path{arXiv:hep-ph/9502213}}, \href
  {https://doi.org/10.1103/PhysRevD.52.271}
  {\path{doi:10.1103/PhysRevD.52.271}}.

\bibitem{Ji:2021mtz}
X.~Ji, {Proton mass decomposition: naturalness and interpretations}, Front.
  Phys. (Beijing) 16~(6) (2021) 64601.
\newblock \href {http://arxiv.org/abs/2102.07830} {\path{arXiv:2102.07830}},
  \href {https://doi.org/10.1007/s11467-021-1065-x}
  {\path{doi:10.1007/s11467-021-1065-x}}.

\bibitem{Lorce:2017xzd}
C.~Lorc\'e, {On the hadron mass decomposition}, Eur. Phys. J. C 78~(2) (2018)
  120.
\newblock \href {http://arxiv.org/abs/1706.05853} {\path{arXiv:1706.05853}},
  \href {https://doi.org/10.1140/epjc/s10052-018-5561-2}
  {\path{doi:10.1140/epjc/s10052-018-5561-2}}.

\bibitem{Lorce:2018egm}
C.~Lorc\'e, H.~Moutarde, A.~P. Trawi\'nski, {Revisiting the mechanical
  properties of the nucleon}, Eur. Phys. J. C 79~(1) (2019) 89.
\newblock \href {http://arxiv.org/abs/1810.09837} {\path{arXiv:1810.09837}},
  \href {https://doi.org/10.1140/epjc/s10052-019-6572-3}
  {\path{doi:10.1140/epjc/s10052-019-6572-3}}.

\bibitem{Diakonov:1987ty}
D.~Diakonov, V.~Y. Petrov, P.~V. Pobylitsa, {A Chiral Theory of Nucleons},
  Nucl. Phys. B 306 (1988) 809.
\newblock \href {https://doi.org/10.1016/0550-3213(88)90443-9}
  {\path{doi:10.1016/0550-3213(88)90443-9}}.

\bibitem{Christov:1995vm}
C.~V. Christov, A.~Blotz, H.-C. Kim, P.~Pobylitsa, T.~Watabe, T.~Meissner,
  E.~Ruiz~Arriola, K.~Goeke, {Baryons as nontopological chiral solitons}, Prog.
  Part. Nucl. Phys. 37 (1996) 91--191.
\newblock \href {http://arxiv.org/abs/hep-ph/9604441}
  {\path{arXiv:hep-ph/9604441}}, \href
  {https://doi.org/10.1016/0146-6410(96)00057-9}
  {\path{doi:10.1016/0146-6410(96)00057-9}}.

\bibitem{Diakonov:1985eg}
D.~Diakonov, V.~Y. Petrov, {A Theory of Light Quarks in the Instanton Vacuum},
  Nucl. Phys. B 272 (1986) 457--489.
\newblock \href {https://doi.org/10.1016/0550-3213(86)90011-8}
  {\path{doi:10.1016/0550-3213(86)90011-8}}.

\bibitem{Goeke:2007fp}
K.~Goeke, J.~Grabis, J.~Ossmann, M.~V. Polyakov, P.~Schweitzer, A.~Silva,
  D.~Urbano, {Nucleon form-factors of the energy momentum tensor in the chiral
  quark-soliton model}, Phys. Rev. D 75 (2007) 094021.
\newblock \href {http://arxiv.org/abs/hep-ph/0702030}
  {\path{arXiv:hep-ph/0702030}}, \href
  {https://doi.org/10.1103/PhysRevD.75.094021}
  {\path{doi:10.1103/PhysRevD.75.094021}}.

\bibitem{Polyakov:2018exb}
M.~V. Polyakov, H.-D. Son, {Nucleon gravitational form factors from instantons:
  forces between quark and gluon subsystems}, JHEP 09 (2018) 156.
\newblock \href {http://arxiv.org/abs/1808.00155} {\path{arXiv:1808.00155}},
  \href {https://doi.org/10.1007/JHEP09(2018)156}
  {\path{doi:10.1007/JHEP09(2018)156}}.

\bibitem{Ji:1996nm}
X.-D. Ji, {Deeply virtual Compton scattering}, Phys. Rev. D 55 (1997)
  7114--7125.
\newblock \href {http://arxiv.org/abs/hep-ph/9609381}
  {\path{arXiv:hep-ph/9609381}}, \href
  {https://doi.org/10.1103/PhysRevD.55.7114}
  {\path{doi:10.1103/PhysRevD.55.7114}}.

\bibitem{Ji:1996ek}
X.-D. Ji, {Gauge-Invariant Decomposition of Nucleon Spin}, Phys. Rev. Lett. 78
  (1997) 610--613.
\newblock \href {http://arxiv.org/abs/hep-ph/9603249}
  {\path{arXiv:hep-ph/9603249}}, \href
  {https://doi.org/10.1103/PhysRevLett.78.610}
  {\path{doi:10.1103/PhysRevLett.78.610}}.

\bibitem{Diehl:2003ny}
M.~Diehl, {Generalized parton distributions}, Phys. Rept. 388 (2003) 41--277.
\newblock \href {http://arxiv.org/abs/hep-ph/0307382}
  {\path{arXiv:hep-ph/0307382}}, \href
  {https://doi.org/10.1016/j.physrep.2003.08.002}
  {\path{doi:10.1016/j.physrep.2003.08.002}}.

\bibitem{Leader:2013jra}
E.~Leader, C.~Lorc\'e, {The angular momentum controversy:
  What\textquoteright{}s it all about and does it matter?}, Phys. Rept. 541~(3)
  (2014) 163--248.
\newblock \href {http://arxiv.org/abs/1309.4235} {\path{arXiv:1309.4235}},
  \href {https://doi.org/10.1016/j.physrep.2014.02.010}
  {\path{doi:10.1016/j.physrep.2014.02.010}}.

\bibitem{Lorce:2017wkb}
C.~Lorc\'e, L.~Mantovani, B.~Pasquini, {Spatial distribution of angular
  momentum inside the nucleon}, Phys. Lett. B 776 (2018) 38--47.
\newblock \href {http://arxiv.org/abs/1704.08557} {\path{arXiv:1704.08557}},
  \href {https://doi.org/10.1016/j.physletb.2017.11.018}
  {\path{doi:10.1016/j.physletb.2017.11.018}}.

\bibitem{Leader:2012ar}
E.~Leader, {A critical assessment of the angular momentum sum rules}, Phys.
  Lett. B 720 (2013) 120--124, [Erratum: Phys.Lett.B 726, 927--927 (2013)].
\newblock \href {http://arxiv.org/abs/1211.3957} {\path{arXiv:1211.3957}},
  \href {https://doi.org/10.1016/j.physletb.2013.01.050}
  {\path{doi:10.1016/j.physletb.2013.01.050}}.

\bibitem{Kim:2020nug}
J.-Y. Kim, H.-C. Kim, M.~V. Polyakov, H.-D. Son, {Strong force fields and
  stabilities of the nucleon and singly heavy baryon $\Sigma_c$}, Phys. Rev. D
  103~(1) (2021) 014015.
\newblock \href {http://arxiv.org/abs/2008.06652} {\path{arXiv:2008.06652}},
  \href {https://doi.org/10.1103/PhysRevD.103.014015}
  {\path{doi:10.1103/PhysRevD.103.014015}}.

\bibitem{Won:2022cyy}
H.-Y. Won, J.-Y. Kim, H.-C. Kim, {Gravitational form factors of the baryon
  octet with flavor SU(3) symmetry breaking}, Phys. Rev. D 106~(11) (2022)
  114009.
\newblock \href {http://arxiv.org/abs/2210.03320} {\path{arXiv:2210.03320}},
  \href {https://doi.org/10.1103/PhysRevD.106.114009}
  {\path{doi:10.1103/PhysRevD.106.114009}}.

\bibitem{Ioffe:1981kw}
B.~L. Ioffe, {Calculation of Baryon Masses in Quantum Chromodynamics}, Nucl.
  Phys. B 188 (1981) 317--341, [Erratum: Nucl.Phys.B 191, 591--592 (1981)].
\newblock \href {https://doi.org/10.1016/0550-3213(81)90259-5}
  {\path{doi:10.1016/0550-3213(81)90259-5}}.

\bibitem{Polyakov:2002yz}
M.~V. Polyakov, {Generalized parton distributions and strong forces inside
  nucleons and nuclei}, Phys. Lett. B 555 (2003) 57--62.
\newblock \href {http://arxiv.org/abs/hep-ph/0210165}
  {\path{arXiv:hep-ph/0210165}}, \href
  {https://doi.org/10.1016/S0370-2693(03)00036-4}
  {\path{doi:10.1016/S0370-2693(03)00036-4}}.

\bibitem{Lorce:2020onh}
C.~Lorc\'e, {Charge Distributions of Moving Nucleons}, Phys. Rev. Lett.
  125~(23) (2020) 232002.
\newblock \href {http://arxiv.org/abs/2007.05318} {\path{arXiv:2007.05318}},
  \href {https://doi.org/10.1103/PhysRevLett.125.232002}
  {\path{doi:10.1103/PhysRevLett.125.232002}}.

\bibitem{Schweitzer:2002nm}
P.~Schweitzer, S.~Boffi, M.~Radici, {Polynomiality of unpolarized off forward
  distribution functions and the D term in the chiral quark soliton model},
  Phys. Rev. D 66 (2002) 114004.
\newblock \href {http://arxiv.org/abs/hep-ph/0207230}
  {\path{arXiv:hep-ph/0207230}}, \href
  {https://doi.org/10.1103/PhysRevD.66.114004}
  {\path{doi:10.1103/PhysRevD.66.114004}}.

\bibitem{Schweitzer:2003ms}
P.~Schweitzer, M.~Colli, S.~Boffi, {Polynomiality of helicity off forward
  distribution functions in the chiral quark soliton model}, Phys. Rev. D 67
  (2003) 114022.
\newblock \href {http://arxiv.org/abs/hep-ph/0303166}
  {\path{arXiv:hep-ph/0303166}}, \href
  {https://doi.org/10.1103/PhysRevD.67.114022}
  {\path{doi:10.1103/PhysRevD.67.114022}}.

\bibitem{Ossmann:2004bp}
J.~Ossmann, M.~V. Polyakov, P.~Schweitzer, D.~Urbano, K.~Goeke, {The
  Generalized parton distribution function (E**u + E**d)(x,xi,t) of the nucleon
  in the chiral quark soliton model}, Phys. Rev. D 71 (2005) 034011.
\newblock \href {http://arxiv.org/abs/hep-ph/0411172}
  {\path{arXiv:hep-ph/0411172}}, \href
  {https://doi.org/10.1103/PhysRevD.71.034011}
  {\path{doi:10.1103/PhysRevD.71.034011}}.

\bibitem{Wakamatsu:2007uc}
M.~Wakamatsu, {On the D-term of the nucleon generalized parton distributions},
  Phys. Lett. B 648 (2007) 181--185.
\newblock \href {http://arxiv.org/abs/hep-ph/0701057}
  {\path{arXiv:hep-ph/0701057}}, \href
  {https://doi.org/10.1016/j.physletb.2007.03.013}
  {\path{doi:10.1016/j.physletb.2007.03.013}}.

\bibitem{LHPC:2007blg}
P.~Hagler, et~al., {Nucleon Generalized Parton Distributions from Full Lattice
  QCD}, Phys. Rev. D 77 (2008) 094502.
\newblock \href {http://arxiv.org/abs/0705.4295} {\path{arXiv:0705.4295}},
  \href {https://doi.org/10.1103/PhysRevD.77.094502}
  {\path{doi:10.1103/PhysRevD.77.094502}}.

\bibitem{Gockeler:2003jfa}
M.~Gockeler, R.~Horsley, D.~Pleiter, P.~E.~L. Rakow, A.~Schafer, G.~Schierholz,
  W.~Schroers, {Generalized parton distributions from lattice QCD}, Phys. Rev.
  Lett. 92 (2004) 042002.
\newblock \href {http://arxiv.org/abs/hep-ph/0304249}
  {\path{arXiv:hep-ph/0304249}}, \href
  {https://doi.org/10.1103/PhysRevLett.92.042002}
  {\path{doi:10.1103/PhysRevLett.92.042002}}.

\bibitem{Kim:1995bq}
H.-C. Kim, M.~V. Polyakov, K.~Goeke, {Nucleon tensor charges in the SU(2)
  chiral quark - soliton model}, Phys. Rev. D 53 (1996) 4715--4718.
\newblock \href {http://arxiv.org/abs/hep-ph/9509283}
  {\path{arXiv:hep-ph/9509283}}, \href
  {https://doi.org/10.1103/PhysRevD.53.R4715}
  {\path{doi:10.1103/PhysRevD.53.R4715}}.

\bibitem{Goeke:2007fq}
K.~Goeke, J.~Grabis, J.~Ossmann, P.~Schweitzer, A.~Silva, D.~Urbano, {The pion
  mass dependence of the nucleon form-factors of the energy momentum tensor in
  the chiral quark-soliton model}, Phys. Rev. C 75 (2007) 055207.
\newblock \href {http://arxiv.org/abs/hep-ph/0702031}
  {\path{arXiv:hep-ph/0702031}}, \href
  {https://doi.org/10.1103/PhysRevC.75.055207}
  {\path{doi:10.1103/PhysRevC.75.055207}}.

\bibitem{Polyakov:2018zvc}
M.~V. Polyakov, P.~Schweitzer, {Forces inside hadrons: pressure, surface
  tension, mechanical radius, and all that}, Int. J. Mod. Phys. A 33~(26)
  (2018) 1830025.
\newblock \href {http://arxiv.org/abs/1805.06596} {\path{arXiv:1805.06596}},
  \href {https://doi.org/10.1142/S0217751X18300259}
  {\path{doi:10.1142/S0217751X18300259}}.

\end{thebibliography}

\end{document}